\documentclass[pra,showpacs,floatfix,twocolumn,10pt]{revtex4}
\usepackage{epsfig,graphicx,amssymb,color,bm,mathbbold,mathrsfs,amsthm}
\usepackage{setspace,amsmath}
\usepackage[colorlinks, linkcolor=blue, urlcolor=blue, citecolor=blue, unicode=true]{hyperref}
\usepackage{url}
\usepackage{booktabs,multirow,makecell}

\begin{document}

%\begin{spacing}{1}
\title{Measurement of the mechanical reservoir spectral density in optomechanical system}

\author{Wen-Zhao Zhang$^{1}$, Xian-Ting Liang$^{1}$, Jiong Cheng$^{1}$\footnote{chengjiong@nbu.edu.cn}, Ling Zhou$^{2}$}
\affiliation{$^{1}$Department of Physics, Ningbo University, Ningbo 315211, China}
\affiliation{$^{2}$School of Physics and Optoelectronic Technology, Dalian University of Technology, Dalian 116024, People's Republic of China}
\date{\today }

\begin{abstract}
To investigate the dynamical behavior of a quantum system embedded in a memory environment,
it is crucial to obtain the knowledge of the reservoir spectral density.
However, such knowledge is usually based on a priori assumptions about the environment.
In this paper, we put forward a method to obtain key information about the reservoir spectral density of an optomechanical resonator without additional assumptions about the spectral shape.
This is achieved by detecting and analysing the optical transmission rate of the emitted light.
In the weak optomechanical single-photon coupling regime, we establish a simple relation between the output light spectrum and the reservoir spectral density.
This provide a straightforward and effective way for reconstructing the spectral density profile in single or even multiple decoherence channels.

\end{abstract}


\maketitle

\section{Introduction}

The system and its surroundings have unavoidable interactions,
therefore realistic quantum systems are essentially open systems \cite{Breuer2002}.
In recent years, the study of the dynamics of open systems becomes active again due to the rapid development
of quantum experimental technologies as well as the possible applications in quantum information science
\cite{Cirac3221,DiVincenzo113,Knill46,Duan032305}.
Meanwhile the features of the non-Markovian quantum process has attracted significant attention in
both theoretical and experimental studies \cite{Reich12430,Chru070406,Xu100502,Liu931,Chin233601,Deffner010402}.
The reliable description of the non-Markovian dynamics depends on the knowledge of the reservoir
as well as the system-reservoir interaction, which may often not be possible to acquire.
Therefore, simplifying assumptions had been made for describing the environment,
which is often attributed to the property of the spectral density \cite{Leggett1987,Breuer2012}.
If the crucial property of the spectral density is known,
then the system dynamics can be fully determined \cite{Breuer2002,Rivas2011}.
In realistic situations, however, due to the interaction with distinct physical environments,
an open system is often affected by different decoherence mechanisms.
In this case, the spectral density function that based on theoretical assumptions,
may not give a reliable description of the environments,
as it is often limited to several classes of functions (i.e., Ohmic or Lorentzian spectral density).

Measuring the spectral density thereby becomes an important task,
and various schemes have been proposed in two-level systems, where the qubit is used as the noise probe,
e.g., pulse sequences \cite{Yuge170504,Alvarez230501,Bylander565},
correlations of single-shot measurement \cite{Fink2013}, waveguide-QED-based measurement \cite{Ciccarello062121},
spontaneous synchronization \cite{Giorgi052121}, open-loop control protocols \cite{Norris150503}, and dynamical evolution detection \cite{Xu032108}.
Recently, the non-Markovian nature of the mechanical heat bath has been revealed experimentally \cite{Groblacher7606},
and the reservoir spectral density is reconstructed by monitoring the mechanical motion with high sensitivity.
This can be achieved in optomechanical systems, which offer an attractive approach to engineer interactions of light and mechanics
that can achieve rather high sensitivity on the displacement measurement via the radiation pressure force \cite{Aspelmeyer1391,Motazedifard023815}.
Optomechanical systems are widely exploited to detect small quantities for sensing purposes,
including gravitational wave detection \cite{Pang124016,Miao211104},
ultra-sensitive force detection \cite{Motazedifard023815,Zhang083022},
adsorbed nano-mass detection \cite{Li141905} and rotating mechanical quantum gyroscopes \cite{Li090311}.
Therefore, with the help of radiation pressure force, the mechanical resonator can hence be used as an ultrasensitive probe measuring the reservoir spectral density.

Due to the experimental progress \cite{Groblacher7606},
the non-Markovian effects of the optomechanical system has been studied extensively \cite{Cheng23678,Zhang063853,Mu012334,Zhang083022,Xiong6053,ZhangWZ063811,Cheng385,Zhao29082,Li1363,Jiang033804}.
The results of these studied are affected by the shape of the spectral density, which is often based on theoretical assumptions.
In this work, we propose a method for measuring the spectral density via the optical transmission rate of the optomechanical system without additional assumptions.
By examining the non-Markovian dynamics of the system related to the output field,
we can reconstruct the spectral density in low frequency regions with high consistency.
For the spectral density interval that far from the mechanical oscillation frequency,
we can still complete the full reconstruction of the spectral density by using the numerical fitting method if the specific form of the spectral density functions is given.
Moreover our method can be also applied to the case when the mechanical resonator has multiple decoherence channels.

%By simulating the detection process, we find that the mechanical probe is sensitive to the spectral density near it's oscillation frequency.

The paper is organized as follows.
In Sec. II, we describe the model and derive the optical transmission rate.
In Sec. III, we demonstrate the application of the scheme with some given examples, including experimental non-Ohmic spectrum, Lorentzian spectrum, Ohmic spectrum,
as well as the case of multiple decoherence channels.
Finally, we conclude in Sec. IV.

\section{Mechanical resonator as the reservoir probe}

We consider a typical cavity optomechanical system consisting of a Fabry-P\'{e}rot cavity with frequency $\omega_{c}$ and a mechanical resonator with frequency $\omega_{m}$.
The mechanical resonator is coupled to a non-Markovian reservoir, which is experimentally achievable \cite{Groblacher7606}.
The corresponding Hamiltonian can be written as $\hat{H}=\hat{H}_{S}+\hat{H}_{EI}$, where
\begin{subequations}
\begin{align}
\hat{H}_{S}&=\hbar\omega_{c}\hat{a}^{\dag}\hat{a}+\frac{\hbar\omega_{m}}{2}(\hat{p}^{2}+\hat{q}^{2}) \notag\\
&~~~-\hbar g_{0}\hat{a}^{\dag}\hat{a}\hat{q}+i\hbar E(e^{-i\omega_{0}t}\hat{a}^{\dag}-e^{i\omega_{0}t}\hat{a}), \label{Hs} \\
\hat{H}_{EI}&=\sum_{l}\frac{\hbar\omega_{l}}{2}[\hat{q}_{l}^{2}+(\hat{p}_{l}-\gamma_{l}\hat{q})^{2}].
\end{align}
\end{subequations}
$\hat{H}_{S}$ is the Hamiltonian of the system \cite{Law2537,Giovannetti023812,Genes033804,Xiong151},
where $\hat{a}^{\dag}$ and $\hat{a}$ are the creation and annihilation operators of the optical mode,
the quadratures $\hat{q}$ and $\hat{p}$ are the dimensionless position and momentum operators of the mechanical mode.
The optomechanical interaction is described by the third term in Eq. \eqref{Hs} with the single-photon coupling coefficient $g_{0}=(\omega_{c}/L)\sqrt{\hbar/2m\omega_{m}}$.
The cavity is driven by a coherent laser with driving strength $E$ and center frequency $\omega_{0}$.
$\hat{H}_{EI}$ is the Hamiltonian of the reservoir as well as the system-reservoir interaction,
which describes a mirror undergoing Brownian motion with the coupling through the reservoir momentum \cite{Giovannetti023812,Ford4419}.
Here, $\omega_{l}$ is the reservoir energy of the mechanical mode, and $\gamma_{l}$ stands for the mirror-reservoir coupling strength.

In the frame rotating at the laser frequency, by eliminating all the environmental degrees of freedom, we can obtain the following Heisenberg-Langevin equations \cite{Cheng23678,Zhang063853}
\begin{subequations}
\begin{align}
\dot{\hat{a}} &=-(i\Delta_{c}+\kappa)\hat{a}+ig_{0}\hat{a}\hat{q}+E+\sqrt{2\kappa}\hat{a}_{in}, \\
\dot{\hat{q}} &=\omega_{m}\hat{p}, \\
\dot{\hat{p}} &=-\Delta_{m}\hat{q}+g_{0}\hat{a}^{\dag}\hat{a}+\int_{0}^{t}d\tau f_{m}(t-\tau)\hat{q}(\tau)+\hat{\xi}_{in}, \label{2c}
\end{align}
\end{subequations}
where $\Delta_{c}=\omega_{c}-\omega_{0}$ is the cavity detuning, and $\Delta_{m}=\omega_{m}+\sum_{l}\omega_{l}\gamma_{l}^{2}$ is the reservoir-induced energy shift.
$\hat{a}_{in}$ is the input noise operator of the cavity, which satisfy the autocorrelation relation $\langle\hat{a}_{in}(t)\hat{a}^{\dag}_{in}(\tau)\rangle=\delta(t-\tau)$.
While $\hat{\xi}_{in}=\sum_{l}\omega_{l}\gamma_{l}[\hat{p}_{l}(0)\cos\omega_{l}t-\hat{q}_{l}(0)\sin\omega_{l}t]$ is the input noise operator of the mechanical resonator.
The non-Markovian effect is fully manifested in the non-local time correlation function $f_{m}(t)=\int\frac{d\omega}{2\pi}J_{m}(\omega)\sin\omega t$,
where $J_{m}(\omega)$ is the mechanical spectral density to be measured.
In the presence of a strong driving field, the intracavity photon number are large enough, thus the quantum fluctuations can be ignored,
we then focus on the classical part that describing the classical phase space orbits of the first moments of operators
\begin{subequations}
\begin{align}
\dot{\alpha} &=-(i\Delta_{c}+\kappa)\alpha +ig_{0}\alpha q+E,  \label{classicala}   \\
\ddot{q} &=\omega_{m}[-\Delta_{m}q+g_{0}|\alpha|^{2}+\int_{0}^{t}d\tau f_{m}(t-\tau)q(\tau)]. \label{classicalq}
\end{align} \label{classical}
\end{subequations}
Eq. \eqref{classicalq} can be formally integrated by utilizing the modified Laplace transformation \cite{Zhang170402}.
For convenience, we set $\omega_{m}=1$, so that Eq. (\ref{classical}) contain only the dimensionless parameters
$g_{0}/\omega_{m}$, $\Delta_{c}/\omega_{m}$, $\Delta_{m}/\omega_{m}$, $\kappa/\omega_{m}$ and $E/\omega_{m}$,
then the solution is
\begin{subequations}
\begin{align}
q(t)&=\dot{Q}(t)q(0)+Q(t)p(0) \notag\\
&~~~+g_{0}\int_{0}^{t}d\tau Q(t-\tau)|\alpha(\tau)|^{2}, \label{solutionq} \\
Q(t)&=\int_{-\infty}^{\infty}\frac{d\omega}{2\pi}\frac{-e^{-i\omega t}}{\omega^{2}-\Delta_{m}-K_{m}(\omega)+\frac{i}{4}\tilde{J}(\omega)}, \label{solutionQ}
\end{align}
\end{subequations}
where the Green's function $Q(t)$ is subjected to the initial conditions $Q(0)=0$ and $\dot{Q}(0)=1$.
$K_{m}(\omega)=\mathcal{P}\int\frac{d\omega^{\prime}}{2\pi}\frac{\omega^{\prime}J_{m}(\omega^{\prime})}{\omega^{2}-\omega^{\prime2}}$
and $\tilde{J}(\omega)=J_{m}(\omega)-J_{m}(-\omega)$ are the real and imaginary part of the Laplace transform of the self-energy correction, respectively \cite{Zhang170402}.
Substituting Eq. \eqref{solutionq} into Eq. \eqref{classicala}, and assuming the mechanical resonator is initially in thermal state, we then have
\begin{align}
\dot{\alpha} &=-(i\Delta_{c}+\kappa)\alpha +E \notag\\
 &~~~+ig_{0}^{2}\alpha(t)\int_{0}^{t}d\tau Q(t-\tau)|\alpha(\tau)|^{2}. \label{Ca}
\end{align}

\begin{figure}
\includegraphics[width=8cm]{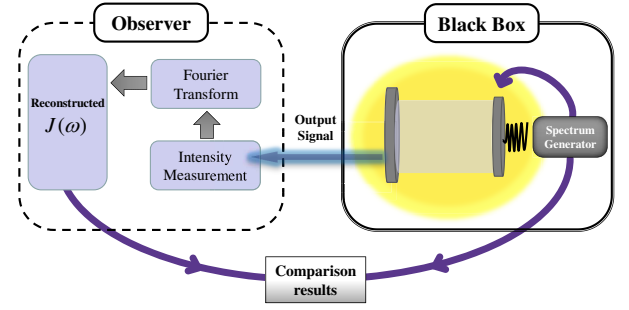}
\caption{
(color online).
Illustration of the probing scheme.} \label{fig1}
\end{figure}

In realistic situations, the experimental realizable cavity optomechanics are still in the single-photon weak coupling limit \cite{Chan89,Teufel359,Kleckner19708,Arcizet71,Groblacher724}.
We thus can perform a first order perturbation solution \cite{Cheng23678} of Eq. \eqref{Ca}, i.e., $\alpha=\alpha_{0}+g_{0}^{2}\alpha_{1}$, where
\begin{subequations}
\begin{align}
\dot{\alpha}_{0}&=-(i\Delta_{c}+\kappa)\alpha_{0}+ E,  \label{Ea0}\\
\dot{\alpha}_{1}&=-(i\Delta_{c}+\kappa)\alpha_{1}  \notag\\
&~~~+i\alpha_{0}(t)\int_{0}^{t}d\tau Q(t-\tau)|\alpha_{0}(\tau)|^{2}. \label{Ea1}
\end{align} \label{Ea}
\end{subequations}
The output field can be obtained by using the input-output relations $\mathscr{E}_{out}+E=2\kappa\alpha(t)$ \cite{Walls1994,Agarwal2010}.
Here we define the modified optical transmission rate $\bar{\eta}=\frac{|\mathscr{E}_{out}|^{2}-E^{2}}{E^{2}}$.
If the cavity field is initially prepared in coherent state with $\langle\hat{a}\rangle=\frac{E}{i\Delta_{c}+\kappa}$.
The transmission rate reduce to $\bar{\eta}=\frac{2\kappa}{E}g_{0}^{2}(A\alpha_{1}+A^{*}\alpha_{1}^{*})$, where $A=\frac{\kappa+i\Delta_{c}}{\kappa-i\Delta_{c}}$.
Utilizing the modified Laplace transformation and together with Eq. \eqref{Ea}, we have (see appendix A)

\begin{figure}[tp]
{\includegraphics[width=8cm]{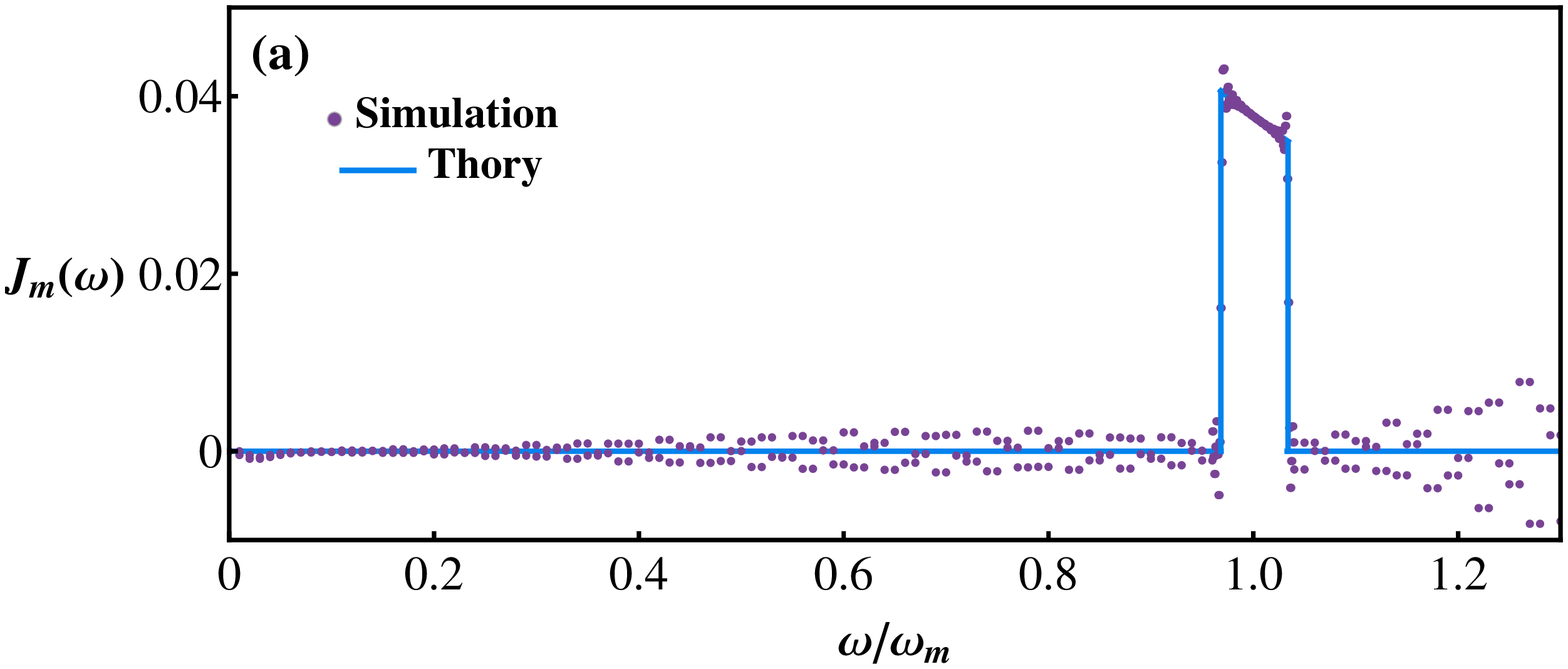}}\\
{~\includegraphics[width=8.2cm]{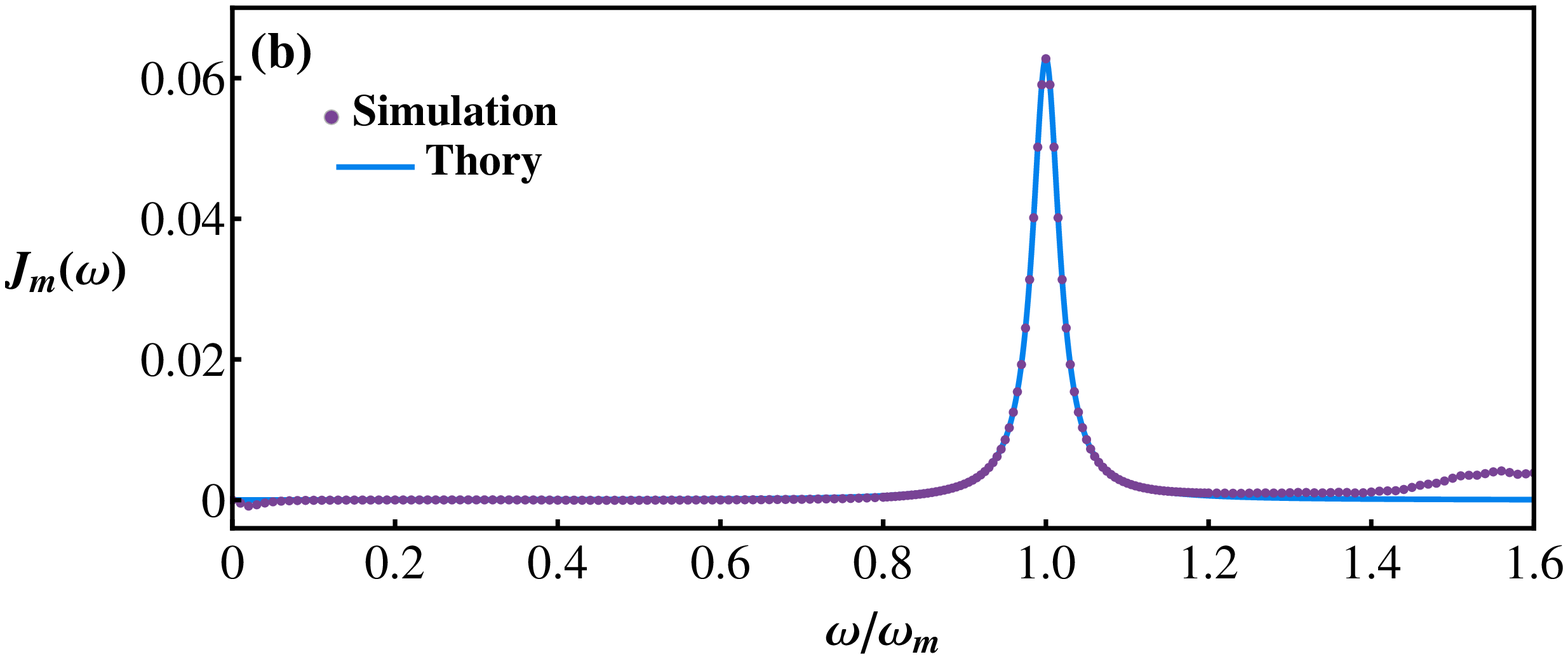}}
\caption{
(color online).
Numerical simulation of the probing scheme.
We compare the given (purple dot) and reconstructed (blue line) spectral density.
In (a), we consider the experimental spectral density reported in Ref. \cite{Groblacher7606}, with $k=-2.3$, $C=6\times10^{-3}$,
while in (b), the Lorentzian spectral density is used and we choose $\Gamma=10^{-2}$, $d=2\times10^{-2}\omega_{m}$.
The other parameters are $E=4\omega_{m}$, $g_{0}=6\times10^{-5}\omega_{m}$, $\Delta_{c}=-7.5\times10^{-3}\omega_{m}$, $\kappa=1.5\times10^{-2}\omega_{m}$.
} \label{fig2}
\end{figure}

\begin{subequations}
\begin{align}
\bar{\eta}(t)&=\frac{1}{\sqrt{2\pi}}\int_{-\infty}^{\infty}d\omega\bar{\eta}(\omega)e^{-i\omega t}, \label{Sluetat} \\
\bar{\eta}(\omega)&=\frac{4\kappa\Delta_{c}g_{0}^{2}E^{2}}{\sqrt{2\pi}(\Delta_{c}^{2}+\kappa^{2})^{2}[\Delta_{c}^{2}-(\omega+i\kappa)^{2}]}  \notag\\
&~~~\times\frac{1}{\omega^{2}-\Delta_{m}-K_{m}(\omega)+\frac{i}{4}\tilde{J}(\omega)}. \label{Sluetaw}
\end{align}  \label{Slueta}
\end{subequations}
Using the solution \eqref{Sluetaw}, we can easily obtain the expression of the spectral density
\begin{align}
\tilde{J}(\omega)=\frac{16\kappa\Delta_{c}g_{0}^{2}E^{2}}{\sqrt{2\pi}(\Delta_{c}^{2}+\kappa^{2})^{2}}
\frac{\mathrm{Im}\{\bar{\eta}(\omega)[(\omega+i\kappa)^{2}-\Delta_{c}^{2}]\}}{|\bar{\eta}(\omega)[(\omega+i\kappa)^{2}-\Delta_{c}^{2}]|^{2}}. \label{Jw}
\end{align}
Apparently, Eq. \eqref{Jw} provide us a straightforward way to extract the spectral density profile from the transmission spectra,
which is easy to record via optical intensity measurements without additional assumptions about the spectral shape.
This is different from the traditional method which detect the undetermined parameters of an assumed spectral structure.
The detected signal, according to Eq. \eqref{Sluetaw}, is proportional to $g_{0}^{2}E^{2}$.
Meanwhile, by solving Eq. \eqref{Ea1}, the applicable condition of the perturbation method can be estimated, i.e., $g_{0}E\ll\sqrt{\Delta_{c}^{2}+\kappa^{2}}$.
This means the accuracy of the solution and the intensity of the signal are constrained by each other.
As a result, our scheme requires the optimization and selection of parameters that balance signal strength and the validity of the derivation.
The strategy of this scheme is based on tracking the system dynamics in time.
The environmental modes that far away from the system center frequency may have negligible impact on the system dynamics due to the rotating-wave approximation.
Eq. \eqref{Sluetaw} describes the dependence of output signal on the spectral density, such dependency reduces rapidly for high-frequency environmental modes.
In other words, the detection effect is inversely proportional to the frequency difference $|\omega-\omega_{m}|$.
To be specific, for $\omega>\omega_{m}$, $\omega^{2}-\Delta_{m}-K_{m}(\omega)+\frac{i}{4}\tilde{J}(\omega)\approx\omega^{2}-\omega_{m}+\frac{i}{4}\tilde{J}(\omega)$,
the ratio $\frac{\tilde{J}(\omega)}{\omega^{2}-\omega_{m}}$ can then be used to evaluate the effectiveness of spectral density detection for specific frequency.

\begin{figure}[tp]
{\includegraphics[width=8.15cm]{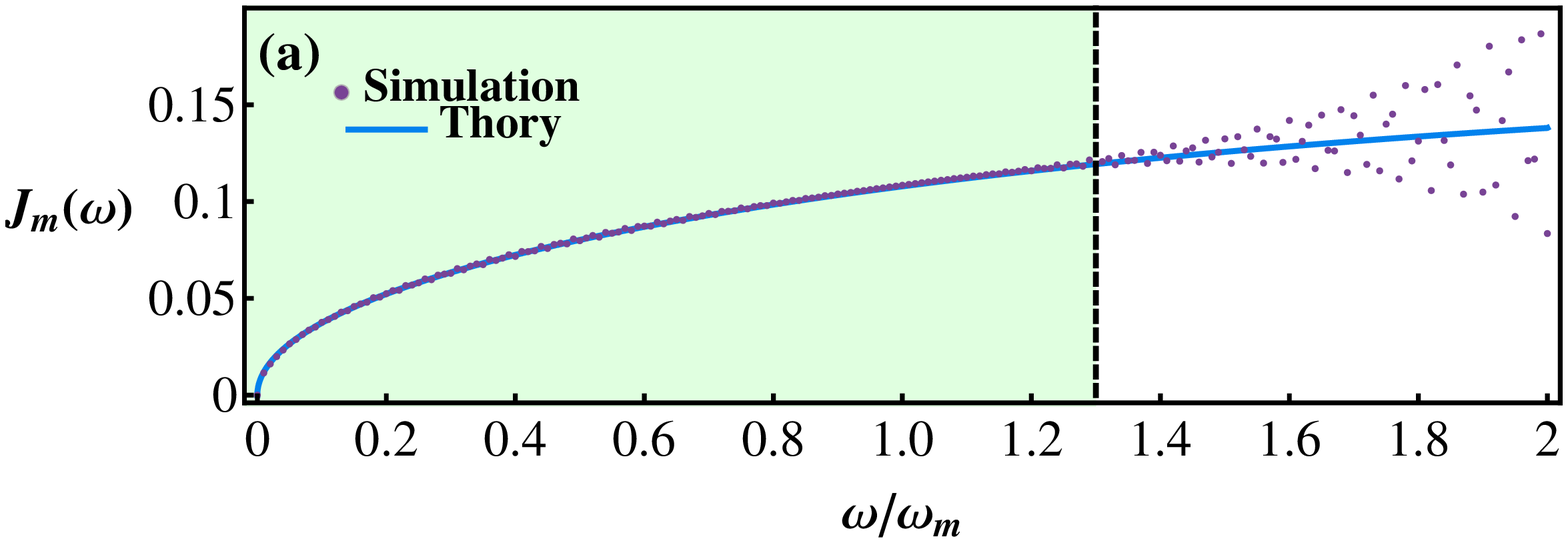}}\\
{\includegraphics[width=8.2cm]{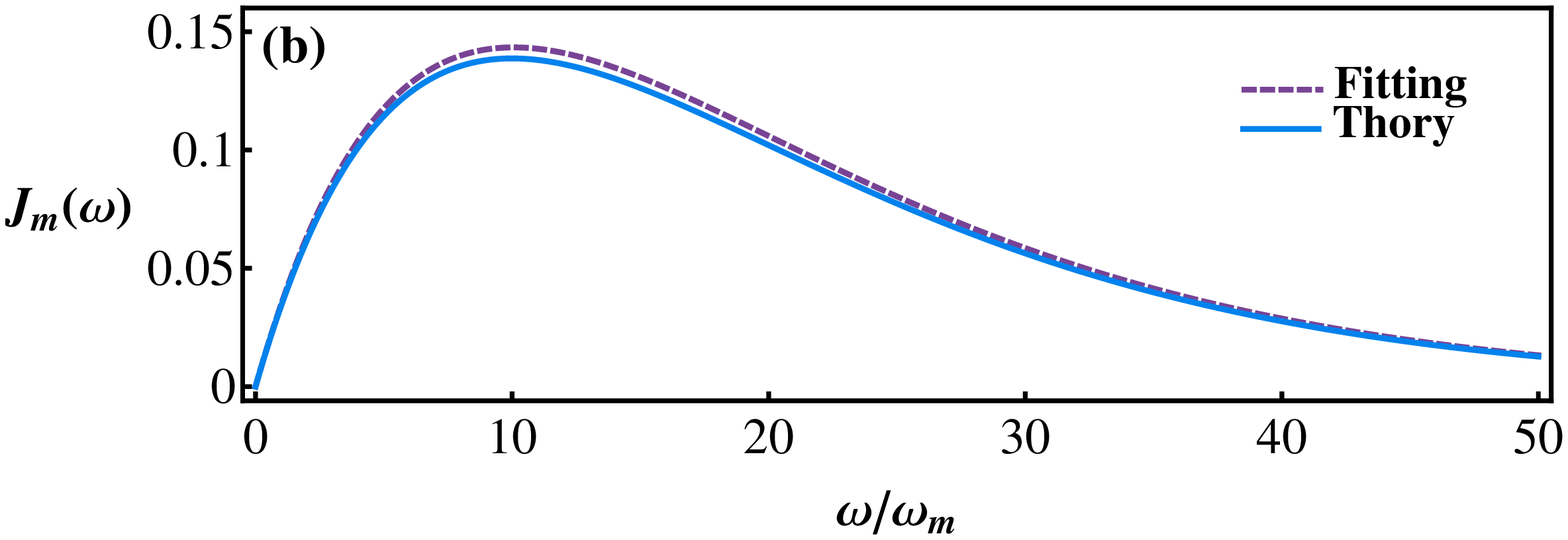}}\\
{\includegraphics[width=8.2cm]{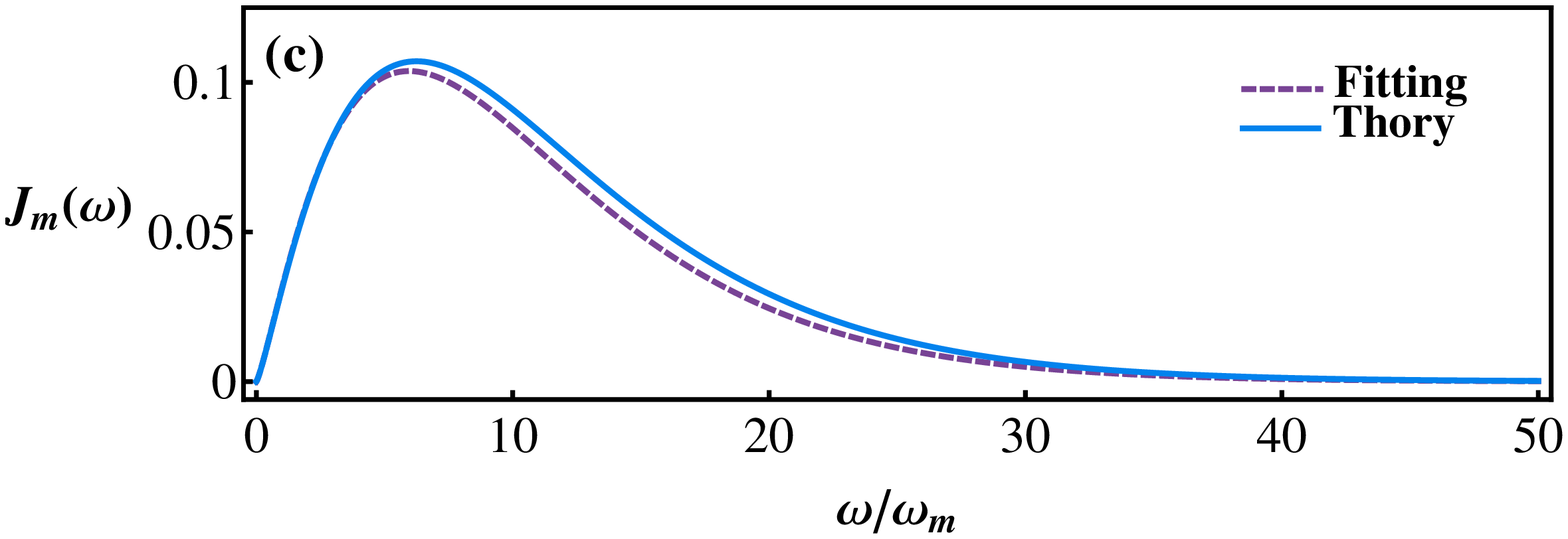}}
\caption{
(color online).
Numerical simulation of the probing scheme with the Ohmic spectrum.
In (a-c), we consider the environment is sub-Ohmicm ($\eta=6\times10^{-3}$, $s=0.5$, $\omega_{c}=10\omega_{m}$), Ohmic ($\eta=6\times10^{-3}$, $s=1$, $\omega_{c}=10\omega_{m}$)
and super-Ohmic ($\eta=9\times10^{-3}$, $s=1.25$, $\omega_{c}=5\omega_{m}$). The other parameters are the same as given in Fig. \ref{fig2}.
} \label{fig3}
\end{figure}

\section{Spectral density reconstruction}

We are now going to discuss the key feature of our scheme.
To simulate the actual experimental process, we use a double-blind method to simulate the system evolution and protocol implementation.
In the actual case, we do not know the dynamics of the system directly, so we use a black box to represent the system.
We assume that, the basic parameters of the system can be detected before we implement the protocol, which has been widely demonstrated in experiments\cite{Aspelmeyer1391}.
As show in Fig. \ref{fig1}, the black box module is used to represent the dynamic evolution of the system in unknown environment.
The spectrum generator is used to generate the environment structure randomly for dynamic simulation.
Our protocol is implemented in the observer module.
In the whole dynamic process, the detector can only get the output signal of the optical field from the black box.
The measurement method is described by four steps.
First, the cavity and mechanical resonator are prepared in coherent and thermal state respectively.
Then, the intensity of the output signal $\bar{\eta}$ is measured through standard intensity measurements.
Third, using the Fourier transform, the detected signal (denote as $S_{\bar{\eta}}$) is transformed to the frequency domain,
i.e., $S_{\bar{\eta}}(\omega)=\frac{1}{\sqrt{2\pi}}\int_{0}^{T}dt S_{\bar{\eta}}(t)e^{i\omega t}$.
Finally, using Eq. \eqref{Jw}, the information of the spectral density is extracted (denote as $S_{\tilde{J}}$).
However, as discussed in the previous section, the intensity of the signal and the accuracy of the solution are constrained by each other.
This set a limit to our detection scheme, but it can still be improved by adding a symmetric detection process (see appendix B).
The reconstructed spectral density can then be written as $R_{\tilde{J}}(\omega)=\frac{S_{\tilde{J}}(\omega,-\Delta_{c})+S_{\tilde{J}}(\omega,\Delta_{c})}{2}$.

\begin{figure}[tp]
{\includegraphics[width=8.2cm]{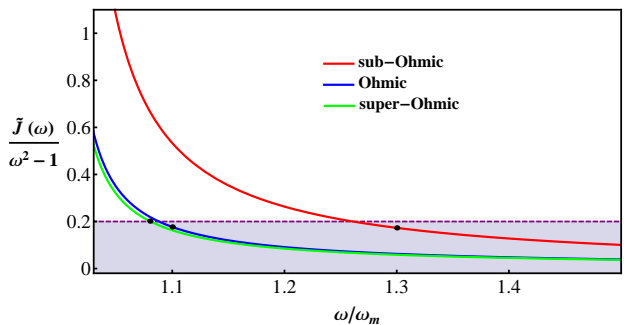}}
\caption{
(color online).
The boundary of the detection scheme for Ohmic-type spectrum. The detection accuracy is negatively correlated to the frequency.
} \label{fig4}
\end{figure}

\subsection{Single Decoherence Channel}

For an experimental demonstration of this scheme, we consider a mechanical resonator with frequency $\omega_{m}=914kHz$.
Experimental study found that, the spectral density is highly non-Ohmic $J_{m}(\omega)=2\pi C\omega^{k}$ and with narrow bandwidth $d\approx0.07\omega_{m}$ \cite{Groblacher7606},
where $C$ is a dimensionless coupling constant.
Due to the rotating wave approximation, the environmental modes that far away from the system central frequency may have negligible impact on the dynamics.
Therefore, we also assume that the spectral density is a local distribution in the vicinity of $\omega_{m}$.
Simulating Eq. \eqref{classical} with the given spectral density, the output field can be determined, and the spectral density can be reconstructed.
Results are plotted in Fig. \ref{fig2}(a), where the calculated spectral density shows a random fluctuation outside the given spectral regions, and it tend to get larger when $\omega>\omega_{m}+\frac{d}{2}$.
This phenomenon could be attributed to the effect of sharp frequency cutoff.
For comparison, in Fig. \ref{fig2}(b) we choose a Lorentzian spectral density with no cutoff, i.e., $J_{m}(\omega)=2\pi\frac{\Gamma d^{2}}{(\omega-\omega_{m})^{2}+d^{2}}$,
but has approximately the same spectral width as the former one. Where $\Gamma$ is the coupling strength and $d$ is the bandwidth of the reservoir spectrum.
The results are obviously improved as expected, but for $\omega\gtrsim1.4\omega_{m}$, the deviation increases evidently.
For spectral density with narrow bandwidth, $\frac{\tilde{J}(\omega)}{\omega^{2}-\omega_{m}}$ decreases rapidly when $\omega>\omega_{m}$.
In this case, our method still works as the main part of the spectral density has been determined.

\begin{figure}[tp]
{\includegraphics[width=8.2cm]{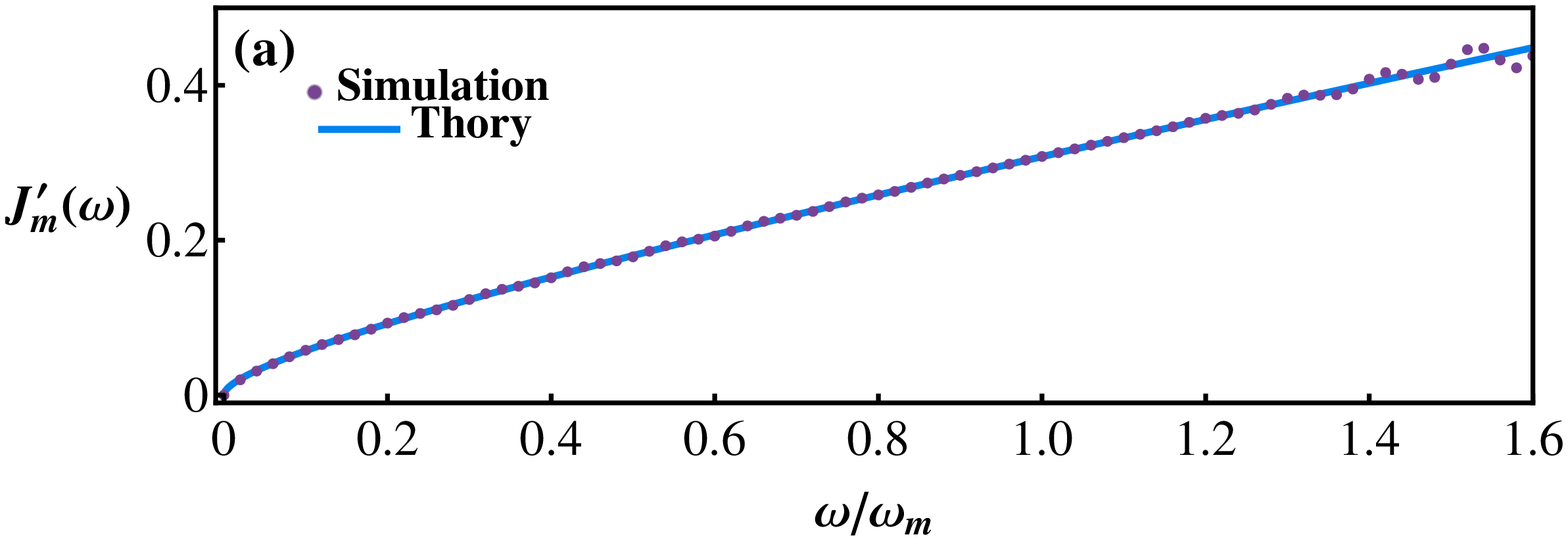}}
{\includegraphics[width=8.2cm]{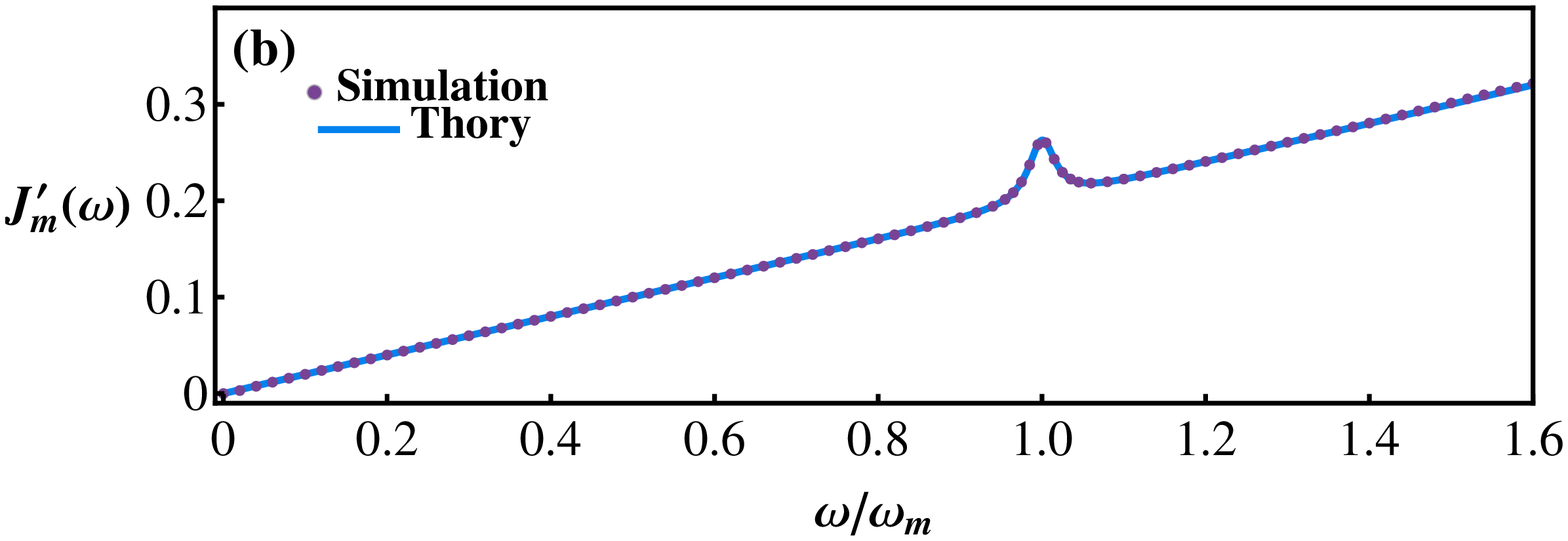}}
{\includegraphics[width=8.2cm]{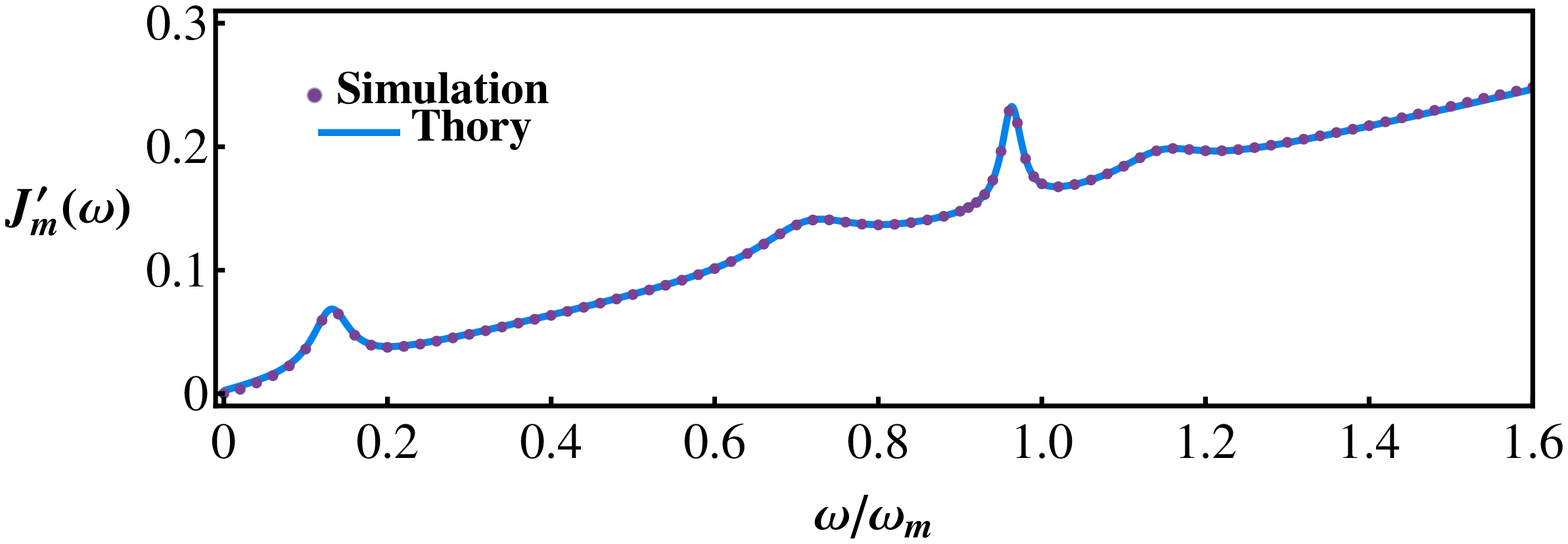}}
\caption{
(color online).
Numerical simulation of the probing scheme when the mechanical resonator is under two decoherence channels:
(a), the non-Markovian damping with the sub-Ohmic spectrum and the Markovian damping.
(b), the non-Markovian damping with the Lorentzian spectrum and the Markovian damping.
(c), the non-Markovian damping with the Lorentzian-like random spectrum and the Markovian damping.
} \label{fig5}
\end{figure}

\renewcommand{\arraystretch}{1.2}
\begin{table*}[htbp]
\centering

\caption{The numerical fitting results: the given parameters (G) and the fitting parameters (F).}   \label{T1}

\begin{tabular}{p{4cm}|p{1cm}p{2cm}p{1cm}p{1.2cm}p{2cm}p{2cm}p{2cm}}

\hline\hline\noalign{\smallskip}

\makecell[c]{Spectral Density}                                & \makecell[c]{G/F} & \makecell[c]{$\eta$}              & \makecell[c]{$s$}     & \makecell[c]{$\omega_{c}/\omega_{m}$} & \makecell[c]{$d/\omega_{m}$}      & \makecell[c]{$\Gamma$}            & \makecell[c]{$\gamma_{m}/\omega_{m}$}   \\

%\noalign{\smallskip}
\hline
%\noalign{\smallskip}

\makecell[c]{\multirow{2}{*}{Ohmic}}                  & \makecell[c]{G}   & \makecell[c]{$6\times10^{-3}$}    & \makecell[c]{$1$}     & \makecell[c]{$10$}                    &                                   &                                   &                                         \\

                                                      & \makecell[c]{F}   & \makecell[c]{$6.21\times10^{-3}$} & \makecell[c]{$1.01$}  & \makecell[c]{$9.99$}                  &                                   &                                   &                                         \\

\hline

\makecell[c]{\multirow{2}{*}{super-Ohmic}}            & \makecell[c]{G}   & \makecell[c]{$9\times10^{-3}$}    & \makecell[c]{$1.25$}  & \makecell[c]{$5$}                     &                                   &                                   &                                         \\

                                                      & \makecell[c]{F}   & \makecell[c]{$9.20\times10^{-3}$} & \makecell[c]{$1.26$}  & \makecell[c]{$4.73$}                  &                                   &                                   &                                         \\

\hline

\makecell[c]{\multirow{2}{*}{sub-Ohmic$+$Markovian}}  & \makecell[c]{G}   & \makecell[c]{$6\times10^{-3}$}    & \makecell[c]{$0.5$}   & \makecell[c]{$10$}                    &                                   &                                   & \makecell[c]{$5\times10^{-2}$}          \\

                                                      & \makecell[c]{F}   & \makecell[c]{$6.54\times10^{-3}$} & \makecell[c]{$0.524$} & \makecell[c]{$10.63$}                 &                                   &                                   & \makecell[c]{$4.83\times10^{-2}$}       \\

\hline

\makecell[c]{\multirow{2}{*}{Lorentzian$+$Markovian}} & \makecell[c]{G}   &                                   &                       &                                       & \makecell[c]{$2\times10^{-2}$}    & \makecell[c]{$10^{-2}$}           & \makecell[c]{$5\times10^{-2}$}          \\

                                                      & \makecell[c]{F}   &                                   &                       &                                       & \makecell[c]{$1.91\times10^{-2}$} & \makecell[c]{$9.93\times10^{-3}$} & \makecell[c]{$5.03\times10^{-2}$}       \\

\noalign{\smallskip}\hline\hline

\end{tabular}

\end{table*}

To test our method in the case of a general bosonic environment, we consider the Ohmic spectrum, $J_{m}(\omega)=2\pi\eta\omega(\frac{\omega}{\omega_{c}})^{s-1}e^{\frac{-\omega}{\omega_{c}}}$.
Where $\eta$ is the coupling strength, $\omega_{c}$ is a high-frequency cutoff.
The parameter $s$ classifies the environment as sub-Ohmic ($0<s<1$), Ohmic ($s=1$), and super-Ohmic ($s>1$), corresponding to Fig. \ref{fig3}(a), (b) and (c).
The given and reconstructed spectral density in Fig. \ref{fig3}(a), agrees well in the light green regions for $\omega<1.3\omega_{m}$, which indicates an unconditional reconstruction region.
When $\omega>1.3\omega_{m}$, the reconstructed curve shows a random oscillation around the theoretical curve, which correspond to a partial reconstruction of the spectral density,
as the specific form of the spectral density functions should be assumed in this case.
Similar results can be also found for Ohmic and super-Ohmic spectrum, they are not present in this paper.
From Fig. \ref{fig2} and Fig. \ref{fig3}(a), we can see that the reconstructed spectral density can be divided into three regions.
The optimal detection region appear at the interval around $\omega_{m}$, which has the highest detecting accuracy.
On the left side is the low frequency region, where our detection scheme is still valid, e.g., see Fig. \ref{fig2} and Fig. \ref{fig3}(a).
However for $\omega>\omega_{m}$, the term $\frac{\tilde{J}(\omega)}{\omega^{2}-\omega_{m}}$ dropped rapidly, followed with the decline of the detecting accuracy.
Now we consider the Ohmic and super-Ohmic spectrum.
Utilizing the explicitly shape of $J_{m}(\omega)$ obtained in low frequency region, we may complete the full reconstruction of $J_{m}(\omega)$ through numerical fitting method.
Fig. \ref{fig3}(b) and (c) show the comparison of the given (solid blue line) with the fitted (dashed purple line) spectral density lines, where a good agreement over the whole range of frequency can be seen.
The detailed parameters are presented in Table \ref{T1}.
In Fig. \ref{fig4}, we mark the points (black points) at which a distinct deviation between simulation and theory occurred for Ohmic-type spectrum.
The unconditional reconstruction region of our detection scheme for Ohmic spectrum therefore can be approximately estimated, i.e., $\frac{\tilde{J}(\omega)}{\omega^{2}-1}\gtrsim0.2$ (above the dashed purple line).
Although in high frequency region, where $\frac{\tilde{J}(\omega)}{\omega^{2}-1}\ll0.2$, our technique is not suitable for the determination of spectral densities with unknown shape.
But obtaining the explicitly shape of $J_{m}(\omega)$ in low frequency region still offer us a way to fully reconstruct the spectral density.

\subsection{Multiple Decoherence Channel}

In realistic situations, the intrinsic decoherence mechanism of the Micro- or nanomechanical resonators is still an unknown issue that been the subject of intense research \cite{Wilson245418,Unterreithmeier027205}.
Apart from the case in which only a single decoherence channel is present,
here we assume that the mechanical resonator is affected by different decoherence mechanisms due to its interaction with distinct physical environments.
To be specific, we first consider the interaction between the resonator and it's reservoir that induces strong non-Markovian features.
Meanwhile the resonator is also exposed to the influences of a completely memoryless channel that could be described by Markovian evolution.
Then Eq. \eqref{2c} is modified by adding a Markovian dissipation term $-\gamma_{m}\hat{p}$.
The general spectrum, in this case, can be expressed as $J^{'}_{m}(\omega)=J_{m}(\omega)+4\gamma_{m}\omega$, where $J_{m}(\omega)$ and $\gamma_{m}$ are unknown quantities to be measured.
The results plotted in Fig. \ref{fig5} show that $J^{'}_{m}(\omega)$ can be also measured with high precision when $\omega<1.5\omega_{m}$.
Meanwhile, numerical fitting is performed by utilizing the measured data if the specific form of the spectral density functions is given.
In Fig. \ref{fig5}(a), we consider the sub-Ohmic spectrum with additional Markovian dissipation.
In Fig. \ref{fig5}(b), we consider the Lorentzian spectrum with additional Markovian dissipation.
The detailed parameters are shown in Table \ref{T1}.
Obviously, the detection accuracy for spectral density with local distribution is higher than the one with wide range of frequencies, even when the mechanical resonator has multiple decoherence channels.

In the example discussing above, the given spectral densities are functions of definite form.
However, in realistic cases, the information about the spectral density is probably unknown.
To simulate this feature in our scheme, we consider a Lorentzian-like random spectrum: $J_{m}(\omega)=2\pi\sum_{i=0}^{4}\frac{\Gamma_{i} d_{i}^{2}}{(\omega-\bar{\omega}_{i})^{2}+d_{i}^{2}}$.
The parameters $\Gamma_{i}$, $d_{i}$ and $\bar{\omega}_{i}$ are randomly generated in our simulation (see appendix C for the detailed parameters).
As shown in Fig. \ref{fig5}(c), the simulated spectrum $J^{'}_{m}(\omega)$ agrees well with the random spectrum.
Thus our scheme can simultaneously determine the unknown spectral density as well as the Markovian dissipation rate with high accuracy.

\section{Conclusion}

We have developed a method for determine the spectral density generated by the environments.
It is based on probing the output field of the cavity, which is affected by the shape of the mechanical spectral density with the help of radiation pressure force.
We have obtained an analytic expression describing the relation between the output light spectrum and the reservoir spectral density,
which enable us to reconstruct spectral densities with unknown shape.
Our method can be easily generalized to the case when the system has multiple decoherence channels.
Furthermore, our results have shown that, the spectral density as well as the dissipation rate can be determined simultaneously.
As the output field detection technology are well established \cite{Langford478,Ourjoumtsev474,Mallet220502,Kapfinger8540}, experimental implementation should be feasible.

%%  The non-Markovian dynamics of the optomechanicl system has been explored theoretically \cite{}. Subsequent studies found that, the non-Markovian effect is helpful for
%%%   The paper is organized as follows. In Sec. II, We
%%% $C=6\times10^{-3}$, $k=-2.3$, $\omega_{min}=885kHz$, $\omega_{max}=945kHz$.
%%%  the spectral density is locally, that is, in the vicinity of an estimate of O

%%%%%%%%%%%%%%%%%%%%%%%%%%%%%%%%%%%%%%%%%%%%%%%%%%%%%%%%%%%%%%%%%%%%%%%%%%%%%%%%%%%%%%%%%%%%%%%%%%%%%%%%%%%%%%%%%%%%%%%%%
\begin{acknowledgments}

We would like to thank Dr. Wen-Lin Li for helpful discussions.
This work is supported by the NSF of China under Grant No. 11704205, No. 11704026, No. 11874099, No. 21773131 and 12074206.
This work is also sponsored by the Natural Science Foundation of Ningbo City (Grant No. 2018A610199) and K.C.Wong Magna Fund in Ningbo University.

\end{acknowledgments}

%%%%%%%%%%%%%%%%%%%%%%%%%%%%%%%%%%%%%%%%%%%%%%%%%%%%%%%%%%%%%%%%%%%%%%%%%%%%%%%%%%%%%%%%%%%%%%%%%%%%%%%%%%%%%%%%%%%%%%%%%%%%%%%
\appendix
%%%%%%%%%%%%%%%%%%%%%%%%%%%%%%%%%%%%%%%%%%%%%%%%%%%%%%%%%%%%%%%%%%%%%%%%%%%%%%%%%%%%%%%%%%%%%%%%%%%%%%%%%%%%%%%%%%%%%%%%%%%%%%%

\section{Derivation of the modified optical transmission rate}

Utilizing the modified Laplace transformation on both side of the optical transmission rate $\bar{\eta}$, the resulting equation is
\begin{align}
\bar{\eta}(z)=\frac{2\kappa}{E}g_{0}^{2}[A\alpha_{1}(z)+A^{*}\alpha_{1}^{*}(z)],
\end{align}
where $\alpha_{1}(z)$ and $\alpha_{1}^{*}(z)$ are the Laplace transformation of $\alpha_{1}(t)$ and $\alpha_{1}^{*}(t)$ respectively, which can be determined through Eq. \eqref{Ea1},
\begin{subequations}
\begin{align}
\alpha_{1}(z)&=\frac{-\alpha_{0}|\alpha_{0}|^{2}}{z-\Delta_{c}+i\kappa}\mathcal{L}[\int_{0}^{t}d\tau Q(\tau)], \\
\alpha_{1}^{*}(z)&=\frac{\alpha_{0}^{*}|\alpha_{0}|^{2}}{z+\Delta_{c}+i\kappa}\mathcal{L}[\int_{0}^{t}d\tau Q(\tau)].
\end{align}
\end{subequations}
Then the optical transmission rate $\bar{\eta}(z)$ reduce to
\begin{align}
\bar{\eta}(z)=\frac{4\kappa\Delta_{c}g_{0}^{2}|\alpha_{0}|^{2}}{(\kappa^{2}+\Delta_{c}^{2})[\Delta_{c}^{2}-(z+i\kappa)^{2}]}\frac{1}{z^{2}-\Delta_{m}-\Sigma_{m}(z)}, \label{A3}
\end{align}
where
\begin{align}
\Sigma_{m}(z)=\int_{0}^{\infty}dtf_{m}(t)e^{izt}=\int\frac{d\omega}{2\pi}\frac{\omega J_{m}(\omega)}{z^{2}-\omega^{2}}, \label{A4}
\end{align}
is the Laplace transform of the self-energy correction \cite{Zhang170402,Cheng23678}.
The modified Bromwich integral for $\bar{\eta}(t)$ is then given by
\begin{align}
\bar{\eta}(t)=\frac{1}{2\pi}\int_{-\infty+i\lambda}^{\infty+i\lambda}dz\bar{\eta}(z)e^{-izt}, \label{A5}
\end{align}
where $\lambda$ is an arbitrary positive real number, therefore the integral is along the upper half plane.
According to Eq. \eqref{A3} and \eqref{A4}, the poles of Eq. \eqref{A5} exist only in the lower half plane.
Thus the integration path can be modified as $\lambda=0$, that is, on the real axis.

\section{The symmetric detection process}

The derivation of the optical transmission rate is based on the perturbation method that only take into account the first order solution of $\alpha$, i.e., $\alpha=\alpha_{0}+g_{0}^{2}\alpha_{1}$.
This method is accurate enough when $g_{0}E/\omega_{m}^{2}\ll1$.
However, on the contrary, the transmission signal is proportional to $g_{0}E/\omega_{m}^{2}$.
In the following, we will prove that the precision of the first order perturbation solution can be further improved by introducing a symmetric detection process.
To be specific, the probe is carried out in two symmetric parameter spaces with $\Delta_{c}$ and $-\Delta_{c}$ respectively.
Here $\Delta_{c}$ is the cavity detuning, which can be easily modulated in experiment.

The error (marked by $\Delta\bar{\eta}$) between the value of $\bar{\eta}$ evaluate in Eq. \eqref{Slueta} and the actual value can be estimated by considering the higher order terms of $\bar{\eta}$ (up to $g_{0}^{4}$),
\begin{eqnarray}
\Delta\bar{\eta}&\approx&\frac{1}{E^{2}}(|\mathscr{E}_{out}^{(2)}|^{2}-|\mathscr{E}_{out}^{(1)}|^{2}) \notag \\
&=&\frac{4\kappa^{2}}{E^{2}}(|\alpha^{(2)}|^{2}-|\alpha^{(1)}|^{2}) \notag \\
&&-\frac{2\kappa}{E}(\alpha^{(2)}+\alpha^{(2)*}-\alpha^{(1)}-\alpha^{(1)*}) \notag \\
&\approx&\frac{4\kappa^{2}}{E^{2}}g_{0}^{4}[|\alpha_{1}|^{2}+\alpha_{0}^{*}\alpha_{2}+\alpha_{2}^{*}\alpha_{0}-\frac{E}{2\kappa}(\alpha_{2}+\alpha_{2}^{*})]. \notag \\
\end{eqnarray}
It is easy to verify the following relations
\begin{eqnarray}
\alpha_{0}(-\Delta_{c})&=&\alpha_{0}^{*}(\Delta_{c}), \notag \\
\alpha_{1}(-\Delta_{c})&=&-\alpha_{1}^{*}(\Delta_{c}), \notag \\
\alpha_{2}(-\Delta_{c})&=&\alpha_{2}^{*}(\Delta_{c}), \notag \\
A(-\Delta_{c})&=&A^{*}(\Delta_{c}).
\end{eqnarray}
Thus it is obvious that
\begin{eqnarray}
\bar{\eta}(-\Delta_{c})&=&-\bar{\eta}(\Delta_{c}), \notag \\
\Delta\bar{\eta}(-\Delta_{c})&=&\Delta\bar{\eta}(\Delta_{c}).
\end{eqnarray}
The detected signal $S_{\bar{\eta}}(\omega)$ can be treat as the sum of the approximate theoretical signal $\bar{\eta}$ and the corresponding error $\Delta\bar{\eta}$.
Using Eq. \eqref{Jw}, we can estimate the error of the reconstructed spectral density which is denoted by $\Delta\tilde{J}$ (for simplicity we set $B=(\omega+i\kappa)^{2}-\Delta_{c}^{2}$).
The error can be written as
\begin{eqnarray}
\Delta\tilde{J}(\omega)&=&\frac{16\kappa\Delta_{c}g_{0}^{2}E^{2}}{\sqrt{2\pi}(\Delta_{c}^{2}+\kappa^{2})^{2}}  \notag \\
&&\times\left[\frac{\mathrm{Im}\{[\bar{\eta}(\omega)+\Delta\bar{\eta}(\omega)]B\}}{|[\bar{\eta}(\omega)+\Delta\bar{\eta}(\omega)]B|^{2}}
-\frac{\mathrm{Im}\{\bar{\eta}(\omega)B\}}{|\bar{\eta}(\omega)B|^{2}}\right] \notag \\
&\approx&\frac{16\kappa\Delta_{c}g_{0}^{2}E^{2}}{\sqrt{2\pi}(\Delta_{c}^{2}+\kappa^{2})^{2}} \frac{\mathrm{Im}\{\Delta\bar{\eta}(\omega)B\}}{|\bar{\eta}(\omega)B|^{2}}.
\end{eqnarray}
If we add a symmetric detection process, then the total error is
\begin{eqnarray}
\Delta\tilde{J}^{\prime}(\omega)&=&\frac{\Delta\tilde{J}(\omega,\Delta_{c})+\Delta\tilde{J}(\omega,-\Delta_{c})}{2} \notag \\
&\approx& \frac{8\kappa\Delta_{c}g_{0}^{2}E^{2}}{\sqrt{2\pi}(\Delta_{c}^{2}+\kappa^{2})^{2}} \notag \\
&&\times \left[ \frac{\mathrm{Im}\{\Delta\bar{\eta}(\omega,\Delta_{c})B\}}{|\bar{\eta}(\omega)B|^{2}}-\frac{\mathrm{Im}\{\Delta\bar{\eta}(\omega,-\Delta_{c})B\}}{|\bar{\eta}(\omega)B|^{2}}\right] \notag \\
&\approx&0.
\end{eqnarray}

Clearly the symmetric detection process can eliminate the effects of higher-order terms, which is actually equivalent to improving the accuracy of the solution up to $g_{0}^{4}$.

\section{The parameters of the random spectrum}

The parameters of the Lorentzian-like random spectral density $J_{m}(\omega)=2\pi\sum_{i=0}^{4}\frac{\Gamma_{i} d_{i}^{2}}{(\omega-\bar{\omega}_{i})^{2}+d_{i}^{2}}$ are shown in Table \ref{T2}.
The other parameters are the same as given in Fig. \ref{fig2}.

\renewcommand{\arraystretch}{1.2}
\begin{table}[htbp]

\caption{The detailed parameters used in Fig. \ref{fig5}(c).}   \label{T2}

\begin{tabular}{p{0.9cm}|p{2.2cm}p{1.4cm}p{2cm}p{1cm}}

\hline\hline\noalign{\smallskip}

                     &  \makecell[c]{$\gamma_{m}/\omega_{m}(10^{-2})$}  & \makecell[c]{$\Gamma_{i}(10^{-3})$}  & \makecell[c]{$d_{i}/\omega_{m}(10^{-3})$}   & \makecell[c]{$\bar{\omega}_{i}/\omega_{m}$}    \\

\hline

\makecell[c]{$i=1$}  &  \makecell[c]{\multirow{4}{*}{$3.843$}}          & \makecell[c]{$7.592$}                & \makecell[c]{$2.776$}                       & \makecell[c]{$0.131$}                          \\

\makecell[c]{$i=2$}  &                                                  & \makecell[c]{$4.701$}                & \makecell[c]{$7.387$}                       & \makecell[c]{$0.714$}                          \\

\makecell[c]{$i=3$}  &                                                  & \makecell[c]{$1.273$}                & \makecell[c]{$1.505$}                       & \makecell[c]{$0.963$}                          \\

\makecell[c]{$i=4$}  &                                                  & \makecell[c]{$3.188$}                & \makecell[c]{$6.229$}                       & \makecell[c]{$1.145$}                          \\

\noalign{\smallskip}\hline\hline

\end{tabular}

\end{table}

%\makecell[c]{\multirow{4}{*}{$3.843\times10^{-3}$}

%%%%%%%%%%%%%%%%%%%%%%%%%%%%%%%%%%%%%%%%%%%%%%%%%%%%%%%%%%%%%%%%%%%%%%%%%%%%%%%%%%%%%%%%%%%%%%%%%%%%%%%%%%%%%%%%%%%%%%%%%

\end{document}